# FORAY-GEN: Automatic Generation of Affine Functions for Memory Optimizations[*]

Ilya Issenin, Nikil Dutt

*Center for Embedded Computer Systems, Donald Bren School of Information and Computer Sciences*
*University of California, Irvine, CA 92697; {isse,dutt}@ics.uci.edu*

## Abstract

*In today's embedded applications a significant portion of energy is spent in the memory subsystem. Several approaches have been proposed to minimize this energy, including the use of scratch pad memories, with many based on static analysis of a program. However, often it is not possible to perform static analysis and optimization of a program's memory access behavior unless the program is specifically written for this purpose. In this paper we introduce the FORAY model of a program that permits aggressive analysis of the application's memory behavior that further enables such optimizations since it consists of 'for' loops and array accesses which are easily analyzable. We present FORAY-GEN: an automated profile-based approach for extraction of the FORAY model from the original program. We also demonstrate how FORAY-GEN enhances applicability of other memory subsystem optimization approaches, resulting in an average of two times increase in the number of memory references that can be analyzed by existing static approaches.*

## 1. Introduction

Customization of memory subsystem in embedded applications is an important part of system design that helps to achieve required power consumption and performance.

Recently a lot of attention has been paid to the use of a scratch pad memory (SPM) in the custom memory subsystems [5][6][7][8][9][10]. Scratch pad memories have lower power consumption than caches [1] and have predictable latency, which is important for real time applications.

Currently, many automated techniques for determining scratch pad configurations rely on compile-time analysis of the program; unfortunately these analysis approaches limit the scope of such memory optimizations, since they assume that the program is written in well structured manner, where all repeating accesses to the memory occur inside *for* loops and furthermore, all memory accesses are generated by ar**ray** references with affine index expressions (which we name the **FORAY** *form*). However, our studies show that many real embedded applications are not written in this form. For example, Figure 1 shows code segments of the *jpeg* application from the MiBench benchmark suite [4], that contain many memory/array access that do not follow the FORAY form*:* not all loop structures are *for* loops, and arrays are accessed using pointers rather than index expressions; furthermore, the iterators of *for* loops are not used in the index expressions of array references. Currently existing memory optimization approaches [5][6][7] are unable to analyze such source codes automatically. This limitation significantly reduces the opportunity for memory optimization and motivates the need for our automated approach.

```
for (ci = 0; ci < cinfo->num_components; ci++)
    for (coefi = 0; coefi < DCTSIZE2; coefi++)
        *last_bitpos_ptr++ = -1;

currow = 0;
while (currow < numrows)
    for (i = rowsperchunk; i > 0; i--) {
        result[currow++] = workspace;
    }
```

**Figure 1. Excerpts from the MiBench benchmark**

We overcome these problems in this paper through two contributions. First, we introduce the *FORAY model* of a program, which is another C program that contains an abstraction of the memory behavior of the original program written in the FORAY form; this FORAY model of the program can be aggressively analyzed and optimized using existing memory subsystem optimization approaches. Second, we present FORAY-GEN, an approach that automatically generates the FORAY model for a given input program, thereby enabling a wider reach for aggressive optimization of the memory subsystem with minimal manual intervention or analysis. Our experimental results show that using FORAY-GEN, we are able to achieve on average a two times increase in the number of analyzable memory references, which in turn greatly increases the scope of memory optimizations.

## 2. Related work

There has been extensive research in using scratch pad memories for storing frequently used data. Some of the approaches [8][9][10] divide all the data used by a program into data objects and profile the application to determine how often and when these data objects are accessed. The most frequently used ones are relocated to scratch pad memory. These techniques do not require any knowledge about the program except declarations of data objects and the information about their use which is obtained by profiling.

[*] This work was partially supported by NSF grants CCR-0203813 and CCR-0205712.

```
for (int i528=0; i528<3; i528++)
    for (int i531=0; i531<64; i531++)
        A[2147447520+4*i531+256*i528]

for (int i1632=0; i1632<1; i1632++)
    for (int i1635=0; i1635<16; i1635++)
        A[268494504+4*i1635]
```

**Figure 2. FORAY models of the programs in Figure 1**

However, such approaches typically do not handle fine-grain array placement, i.e., determining and placing the most frequently used parts of the arrays in the scratch pad memory in order to improve power and performance.

The works [5][6][7] are complementary to the above techniques and are focused on optimization of array accesses. To determine the parts of the array that are going to be reused and which can be placed in the scratch pad memory, a static analysis of the program is performed. This requires all memory accesses (that need to be analyzed) to be expressed as array accesses with affine index expressions (i.e., the FORAY form). Using static analysis ensures that selected array references always expose expected behavior regardless of the input data. Unfortunately, application developers write code in styles suited to individual tastes, and therefore do not always express memory references in the FORAY form. Indeed our studies show that existing benchmarks and source codes contain many memory references that are not written in the FORAY form. This either limits the scope of these scratch pad memories optimization techniques, or requires system designers to manually transform the memory references in the input code to follow the FORAY form. Such manual transformation is cumbersome, time-consuming, and prone to errors; it rapidly becomes infeasible for large applications. For example, the *jpeg* application [4] has more than 30000 lines of code with hundreds of loops. Limiting the scope of manual transformations may result in many missed opportunities.

The work of Franke et al. [3] addresses the problem of converting pointer accesses to array accesses with explicit index functions by performing compile-time analysis of the code. However, since static pointer analysis without any restrictions is intractable, many assumptions about the program structure are made.

Our FORAY-GEN approach automatically generates code in the FORAY form by performing additional dynamic analysis that exposes memory references in the FORAY form. Indeed, we are able to discover all memory references that actually behave as array references with affine index expressions but expressed in any other way in a source program.

## 3. Design flow using FORAY-GEN

We now outline the use of FORAY-GEN in the context of typical memory subsystem optimization techniques.

First, we define the FORAY model: it is a C program consisting of any combination of *for* loops and ar**ray** references (**FORAY**), with all array index expressions being affine functions of outer loop iterators. Array references can also have *partial affine index expressions*, which is described in more detail is Section 4. Note that the FORAY model captures the memory behavior of *only* those memory references in the original program whose access addresses can be described by affine function of outer loop iterators. All other accesses or program functionality are left out of the model. Thus the FORAY model is not functionally equivalent to the original program, but is an abstraction enabling aggressive memory analysis that faithfully captures all memory references that are amenable to optimization by current memory optimization approaches; it is exclusively used for enhancing the reach of such optimizations.

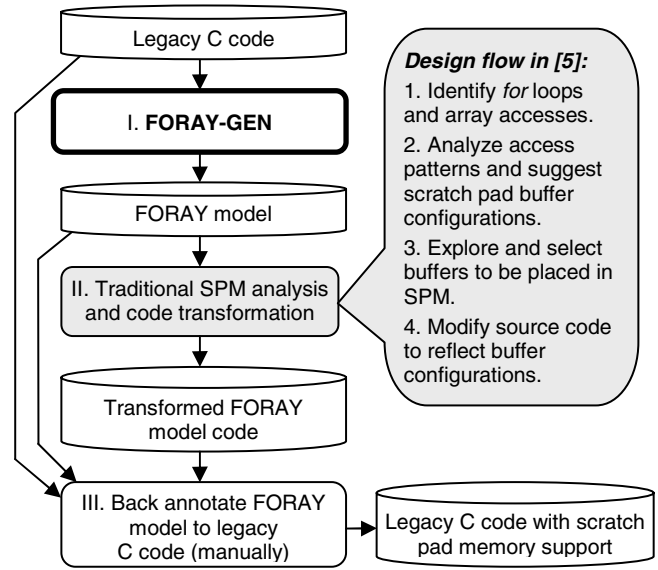

**Figure 3. The use of FORAY model in memory subsystem design**

The FORAY models for the examples in Figure 1 are shown in Figure 2, where the memory access behavior is restructured into well structured *for* loops containing array references with affine index expressions.

The use of the FORAY model in the context of typical SPM subsystem optimization techniques is shown in Figure 3. In Phase I, FORAY-GEN takes legacy C code as input and generates a FORAY model of the application. In Phase II, this FORAY model is optimized using a typical approach for designing a scratch pad memory subsystem (expanded in the shaded call-out of Figure 3). These SPM optimizations typically scan the memory accesses (in FORAY form) and perform compile-time (static) analysis of which data is reused and can be placed to the buffers in the scratch pad memory. Several buffer configurations are suggested and one of them is selected during the design space exploration. The output of Phase II contains FORAY model source code that is changed to access the scratch pad memory and perform the necessary data transfers between scratch pad buffers and main memory Finally in Phase III, (i.e., after determining which arrays benefit from being placed in the scratch pad memory), the designer can manually back-annotate the modified FORAY model code into the original program. At the end of this entire flow, the legacy C code is transformed into C code containing aggressive scratch pad memory optimizations.

Note that the amount of back-annotation required by the designer is *significantly less* than the efforts needed to manually convert the whole original program to FORAY form, since only a few arrays are typically selected to be placed in scratch pad memory.

| (a) Original program | (b) Annotated program | (c) Trace file | Checkpoint: 15 |
|---|---|---|---|
| char q[10000];<br>char *ptr = q;<br>int i, t1 = 98;<br>while (t1 < 100) {<br>   t1++;<br>   ptr += 100;<br>   for (i=40; i>37; i--) {<br>      *ptr++ = i*i % 256;<br>   }<br>} | char q[10000]; char *ptr = q;<br>int i, t1 = 98;<br>{ CHECKPOINT(12); while (t1 < 100) {<br>  CHECKPOINT(13); {<br>    t1++; ptr += 100;<br>    {CHECKPOINT(15); for (i=40; i>37; i--)<br>    {CHECKPOINT(16); {<br>      *ptr++ = i*i % 256;<br>    } CHECKPOINT(14); }}<br>} CHECKPOINT(17); }} | Checkpoint: 12<br>Checkpoint: 13<br>Checkpoint: 15<br>Checkpoint: 16<br>Instr: 4002a0 addr: 7fff5934 wr<br>Checkpoint: 14<br>Checkpoint: 16<br>Instr: 4002a0 addr: 7fff5935 wr<br>Checkpoint: 14<br>Checkpoint: 16<br>Instr: 4002a0 addr: 7fff5936 wr<br>Checkpoint: 14<br>Checkpoint: 17<br>Checkpoint: 13 | Checkpoint: 16<br>Instr: 4002a0 addr: 7fff599b wr<br>Checkpoint: 14<br>Checkpoint: 16<br>Instr: 4002a0 addr: 7fff599c wr<br>Checkpoint: 14<br>Checkpoint: 16<br>Instr: 4002a0 addr: 7fff599d wr<br>Checkpoint: 14<br>Checkpoint: 17 |
| (d) FORAY model<br>for (int i12=0; i12<2; i12++)<br> for (int i15=0; i15<3; i15++)<br>   A4002a0[2147440948+1*i15+103*i12] | | | |

**Figure 4. Example of FORAY-GEN model extraction**

The main contribution of this paper is the automation of Phase I (FORAY-GEN) in this flow. Note that in Phase II, the generated FORAY model can be used by any scratch pad (such as [5][6][7]) or other memory optimization techniques.

## 4. FORAY-GEN

We now describe our profile-based FORAY-GEN approach, as outlined in Algorithm 1 (Figure 5).

*Algorithm 1: Outline of **FORAY-GEN** model extraction*
1. Annotate the program
2. Profile the program
3. Analyze the trace file and
   3.1. Reconstruct loop structure from the trace (Algorithm 2)
   3.2. Determine affine index expression for each memory reference (Algorithm 3)
4. Purge uninteresting memory references

**Figure 5. Algorithm outline of FORAY-GEN**

First, in Step 1 of Algorithm 1 we annotate the loop structures (*for*, *do* and *while* loops) in the source code of a program with checkpoints. Each checkpoint statement is converted later into special assembler instructions that instruct the simulator to add information about execution of the checkpoint to the trace file. An example of an original and the annotated program is shown in Figure 4(a) and (b).

In Step 2 of Algorithm 1 we profile the compiled program by running it on an instruction set simulator. The simulator records the information about each memory access (instruction address and address of access) as well as the information about all checkpoint instruction executed in the trace file. The trace file for the program in Figure 4(a) is shown in Figure 4(c).

In Step 3 of Algorithm 1 we process the trace file generated in the previous step and reconstruct the program loop structure (Step 3.1). In addition to that, during the same pass on the trace file, for each memory reference we try to find an affine expression that can describe all the access addresses of this reference (Step 3.2). These steps are discussed in more detail later in this section.

In Step 4 of Algorithm 1 we use a heuristic that filters out all memory references that cannot be effectively predicted by an affine function. We leave only the references that

- have affine index expression that includes at least one iterator (by this condition we exclude all references that do not have regular access patterns);
- have been executed no less than $N_{exec}$ times;
- address at least $N_{loc}$ different memory locations.

The constants $N_{exec}$ and $N_{loc}$ are selected to leave only references that may benefit from being placed in the scratch pad memory by array-oriented design techniques. In our experiments we used the values of 20 and 10 correspondingly to eliminate small arrays that can fit in the scratch pad completely (and thus can be handled by other techniques like [8][9][10] with less overhead) and to eliminate references which do not exhibit a lot of reuse.

The resulting FORAY model for the program in Figure 4(a) is shown in Figure 4(d).

Next, we describe our implementation of the trace file analysis (Steps 3.1 and 3.2 of FORAY-GEN in Algorithm 1).

**Reconstructing loop structure from trace (Step 3.1 of Algorithm 1)**

Figure 6 outlines the algorithm we developed that reconstructs the loop/reference structure of the original program from the trace file (Step 3.1 of the Algorithm 1).

*Algorithm 2: Building the loop structure of a program*
**Input:** *trace file*
**Output:** *a tree with loop and memory reference nodes*
1. Read next record from the trace file.
2. If it is memory access information, call Algorithm 3 (see below).
3. If it is checkpoint, depending of the type of checkpoint (beginning-of-the-loop, beginning-of-the-loop-body or end-of-the-loop-body checkpoint), move current loop node pointer up or down in the tree or create a new loop node.
4. Go to Step 1.

**Figure 6. Algorithm outline for loop structure reconstruction**

The loop/reference structure of the program is represented as a tree with loop and memory reference nodes. During the processing of the trace file each loop node also maintains the current value of a variable that counts the number of loop iterations. The values of these iterators are used for determining index expressions in Step 3.2 of the Algorithm 1.

**Identifying affine index expressions for memory references (Step 3.2 of Algorithm 1)**

We first give an example of the analysis we perform, and then describe our algorithm (Algorithm 3) for identifying

```
int foo() {                          int foo(int offset) {
    int ret = 0;                         int ret = 0;
    int A[100];                          …
    …                                    for (i=0; i<10; i++)
    for (i=0; i<10; i++)                     for (j=0; j<10; j++)
        for (j=0; j<10; j++)                     ret += A[j+10i+offset];
            ret += A[j+10i];             return ret;
    return ret;                      }
}
int main() {                         int main() {
    …                                    …
    for (x=0; x<10; x++)                 for (x=0; x<10; x++)
        for (y=0; y<10; j++) {               …
            …                                tmp += foo(lines[x]);
            tmp += foo();                }
        }                            }
}
```

**Figure 7. Examples when access addresses can not be described by one affine function**

affine index expressions.

Affine address functions have the following form:

$index = const + C_1*iter_1 + C_2*iter_2 + … + C_N*iter_N,$

where *index* is the access address of the memory reference; *const* – is some constant term (base memory address); *N* is the loop nest level of the memory reference; $C_1..C_N$ are integer coefficients; and $iter_1 .. iter_N$ are current values of the loop iterators ($iter_1$ is the iterator of the innermost loop).

In the example of Figure 4a the exact index expression for all memory references can be found. However, often it is not possible to describe access addresses by one affine function. Two examples of such cases are shown in Figure 7. In the first case, for each call to the function *foo()*, a local array *A[]* may be allocated to a different memory location. In the second case, the globally defined array *A[]* is not reallocated, but the offset is a data-dependent parameter that is passed to the function *foo()*. In both instances, the access addresses within the function *foo()* are regular and can be described by one affine function. However, with every iteration of the outer loops *x* and *y,* the constant term in the index expression changes in an unpredictable manner. In these cases our analysis algorithm finds a *partial affine index expression* for the loop iterators for which the value of index can be predicted by the expression in the form of

$index = const(iter_{M+1}.. iter_N) + C_1*iter_1 + … + C_M*iter_M ,$

where *M < N*; *const* changes every time with the increment of $iter_{M+1}.. iter_N$.

Finding memory references that are expressible as *partial affine index expression* enhances existing SPM analysis approaches [5][6][7]. Indeed, when no full affine index expression exist, *SPM approaches can still perform analysis on a limited number of loops* (which are inside the function *foo()* in the example in Figure 7) as if no other outer loops existed and suggest possible scratch pad buffer configurations that hold the data reused in loops in function *foo()*.

Algorithm 3 in Figure 8 describes how index expressions are determined; it is called for each memory access recorded in the trace file in Step 2 of Algorithm 2.

In Step 1 of Algorithm 3 we create a new memory reference node if we have not encountered this memory reference in the current position inside the loop tree.

***Algorithm 3:*** *Finding array index expressions*
**Input:** *memory access information from the trace file; loop tree and location of the memory reference in it (current loop node); current values of the loop iterators (from Algorithm 2)*
**Output:** *full or partial index expr. for each array reference*
**Variables:** *each memory reference node keeps the following information:*
*N - loop nest level of the current loop node;*
*M – number of iterators included in partial index expression;*
*CONST – constant term of the index expression;*
$C_1..C_N$ *– integer coefficients of the iterators in the affine index expression or UNKNOWN*
$IT_1..IT_N$ *- current values of the iterators (from Algorithm 2)*
$ITP_1..ITP_N$ *- values of the iterators when the same reference was executed the previous time*
$S_1..S_N$ *– a vector of binary values; used for determining the number of iterators in the partial affine index expression*
*IND – address of the current access*
*INDP – address of the previous access*

**Begin**
 1. Search if current memory reference has been already added to the current loop node (see algorithm 2). If not,
    • add the reference node to the current loop node;
    • set CONST = IND; M = N;
    • for i = 1..N, set $C_i$ = UNKNOWN; $S_i$ = 0;
    • go to Step 7.
    If yes, continue.
 2. Calculate H = size of the set HS = {[i]: i = 1..N ∩ $IT_i$ ≠ $ITP_i$ ∩ $C_i$ = UNKNOWN}; k ∈ HS.
 3. if (H = 1) Calculate ADJ = $\sum_{\substack{i=1..N, IT_i \neq ITP_i, \\ C_i \neq UNKNOWN}} IT_i * C_i$ ;
    Calculate $C_k$ = (IND – ADJ – INDP) / ($IT_k$ – $ITP_k$).
 4. if (H > 1) Mark current reference as non-analyzable.
 5. Calculate INDC = CONST + $\sum_{\substack{i=1..N \\ C_i \neq UNKNOWN}} IT_i * C_i$ .
 6. if (INDC ≠ IND)  // prediction was wrong; recalculate CONST
    for i = 1..N
        if ($IT_i$ = $ITP_i$) $S_i$ = 1;
    Calculate CONST = CONST + IND – INDC;
    M = 0; // adjust the number of iterators in the partial index expr.
    for i = 1..N
        if ($S_i$ = 0) M = i-1;
 7. Return to processing of the next statement in the trace file.
**End**

**Figure 8. Algorithm for finding index expressions**

If the current memory reference has been encountered before, in Step 2 of Algorithm 3 we calculate the number *H* of coefficients that have an *UNKNOWN* value and whose corresponding iterators have changed their value after the previous execution of the same memory reference. In variable *k* we save the number of one of such iterators.

In Step 3 of Algorithm 3 we calculate the value of an unknown coefficient if only one iterator with unknown coefficient has changed its value.

If there are several iterators with unknown coefficients that have changed their values, in Step 4 of the Algorithm 3 we mark the current reference as non-analyzable and exclude it from further consideration. Our experiments show that there are very few such references in the benchmarks we studied.

In Step 5 of Algorithm 3 we calculate the predicted value of affine index expression.

In Step 6 of Algorithm 3 we check if this predicted value matches actual address of the memory access. If the prediction is wrong, we mark all iterators that have not changed their values, recalculate the new value of the constant term in the affine index expression and update the value of M that specifies the number of iterators included in the partial index expression. This is done by finding the outermost iterator that has changed its value in every misprediction case. All innermost iterators up to that one are forming a partial affine index expression.

Note that during the building of the FORAY model of a program, each record in the trace file is accessed only once and the records are accessed in the order the information is written to the file. This means that the proposed algorithm can be executed during profiling (Step 2 of the Algorithm 1) and there is no need to save the (typically large) trace file. In this case the space complexity is constant with respect to the number of instructions profiled.

The computational complexity of our approach is dominated by the number of calls of the Algorithms 2 and 3, and is linear with respect to the number of profiled instructions. Since the maximum loop nest level is limited in real programs, the complexity of the Algorithms 2 and 3 is constant on average if we use hash tables for the searches in Algorithm 2 and in Step 1 of Algorithm 3.

**Inter-function optimizations**

Our approach also helps a designer by providing hints on which functions should be inlined to enable more aggressive memory optimizations – this is particularly helpful when a designer attempts to manually restructure code for improved performance and power.

The FORAY model does not have any special representation of the hierarchical structure of a program, which is why functions appear to be inlined in our model. If FORAY-GEN inlines a function in more than one place, this provides a hint to programmer that it may be beneficial to duplicate the function in the original code. In the example shown in Figure 9 the function *foo()* is called in two different places. In those instances the access pattern to array *A[]* inside the body of the function is different. Optimizing the function for the first access pattern can produce suboptimal results for the case when *foo()* is called in the loop *y*. Duplicating the function as suggested by our approach produces better results by allowing each access pattern to be optimized separately.

## 5. Experimental results

We now describe the implementation of the FORAY-GEN framework presented in Algorithm 1. Our parser for Step 1 of Algorithm 1 parses source C files of a benchmark and inserts checkpoint instructions. We use the SimpleScalar functional simulator [2] for profiling (Step 2 of Algorithm 1). The simulator was modified to process checkpoint instructions and to generate the trace file. Steps 3 and 4 of Algorithm 1 are implemented in a separate program that reads the trace file, performs analysis and extracts the FORAY model.

To evaluate our approach, we used several benchmarks from the MiBench suite [4] that contains a set of embedded codes written in standard "C". The following benchmarks were

```
int main() {
    …
    for (x=0; x<10; x++) {
        …
        tmp += foo(10*x);
    }
    for (y=0; y<20; y++) {
        …
        tmp += foo(2*y);
    }
}

int foo(int offset) {
    for (i=0; i<10; i++)
        ret += A[i+offset];
    return ret;
}
```

**Figure 9. Example of a code where function inlining may be beneficial**

used: *jpeg* (image compression), *lame* (MP3 encoder), *susan* (image recognition program), *fft* (Fast Fourier Transformation), and the *gsm* and *adpcm* encoders.

### 5.1. Advantages of using FORAY-GEN

In this section we present the benefits of using our automated FORAY-GEN tool to extract FORAY models for the benchmarks.

First, we estimated the complexity of manual analysis needed to extract FORAY model that can be automated by our approach. The number of lines of code in each benchmark and total number of loops (excluding the loops that were not executed during profiling) are presented in Table I. Breakdown of the loops by the type (*for*, *while* or *do* loops) is also presented.

Table I shows that some of the typical embedded applications are fairly large and have hundreds of loops. Manual analysis of every loop in such applications is error-prone and impractical. We can also conclude that although *for* loops are used predominantly in the programs, there are a lot of other loops (23% on average) that should not be ignored during the analysis (as described earlier, existing approaches for loop-based program analysis consider only *for* loops).

We have also estimated how many of the useful references for scratch pad memory analysis [5][6][7] are not in the FORAY form in the original program and thus can not be analyzed by their static approaches. Table II shows the number of loops and references that can be expressed in FORAY form (those that were found by our algorithm and included in the FORAY model), and percentage of loops and references that are not in FORAY form in the original program.

As we can see from the Table II, in all the benchmarks except *fft* (the shortest one), a considerable number of loops and memory references (64% and 60% on average) are not in FORAY form and thus cannot be statically analyzed by traditional scratch pad memory optimization techniques without using our approach.

### 5.2. Memory behavior of the FORAY model

We now analyze the effectiveness of our extracted FORAY model in capturing the memory behavior of the original program.

We analyzed three aspects of memory behavior: the number of memory references, the number of accesses these references make, and the footprint (the number of different addresses accessed). Recall that the FORAY model captures only the relevant memory references in the original program for SPM optimizations. All memory references in the original program can be divided into three categories: references captured by the FORAY model, system library memory references (not handled by FORAY-GEN), and all other memory references that can not be represented by the FORAY model. Note that system libraries are specific for the platform the code is compiled on and therefore system libraries we used are not relevant to real embedded systems.

**Table I. Benchmark complexity and loop distribution**

| Benchmark name | Number of lines | Total number of loops | Number of *for* loops | Number of *while* loops | Number of *do* loops |
|---|---|---|---|---|---|
| jpeg | 34590 | 169 | 65% | 34% | 1% |
| lame | 22846 | 479 | 83% | 8% | 9% |
| susan | 2173 | 14 | 79% | 21% | 0% |
| fft | 493 | 11 | 100% | 0% | 0% |
| gsm | 7089 | 38 | 87% | 13% | 0% |
| adpcm | 782 | 2 | 50% | 50% | 0% |

**Table II. Loops and references converted into FORAY form by our approach**

| Benchmark name | Loops and references that can be represented by FORAY form (returned by Algorithm 1) | | Percentage of loops and references that are not in FORAY form in the original program (out of those that can be expressed in FORAY form) | |
|---|---|---|---|---|
| | Number of loops | Number of references | Number of loops | Number of references |
| jpeg | 73 | 73 | 41% | 38% |
| lame | 232 | 980 | 42% | 38% |
| susan | 9 | 10 | 78% | 50% |
| fft | 8 | 19 | 0% | 0% |
| gsm | 17 | 86 | 59% | 74% |
| adpcm | 2 | 1 | 100% | 100% |

**Table III. Memory behavior of the FORAY models**

| Benchmark name | Total number | | | Included in FORAY model | | | In system calls | | | Other |
|---|---|---|---|---|---|---|---|---|---|---|
| | References | Accesses | Footprint | Ref. | Accesses | Footprint | Ref. | Accesses | Footprint | Footprint |
| jpeg | 6151 | 8.3M | 123625 | 1% | 27% | 87% | 33% | 2% | 9% | 91% |
| lame | 16805 | 43M | 127052 | 6% | 22% | 26% | 40% | 20% | 33% | 66% |
| susan | 1162 | 5.0M | 24778 | 1% | 66% | 72% | 85% | 1% | 47% | 1% |
| fft | 2420 | 22M | 28804 | 1% | 1% | 57% | 95% | 96% | 43% | 29% |
| gsm | 2091 | 37M | 16215 | 4% | 32% | 5% | 49% | 3% | 93% | 8% |
| adpcm | 546 | 5.5M | 4964 | 0.2% | 28% | 20% | 97% | 0.2% | 68% | 12% |

As we can see from the breakdown in Table III, although very few references in the source code are described by the FORAY model (and worth considering for analysis by techniques [5][6][7]) – 2.2% on average – these references make up to 66% (29% on average) of all accesses and cover up to 87% (44% on average) of the address space.

It is important to note that the data shown in the table includes memory references that are not present explicitly in the source C code (e.g., those placing arguments to the stack before performing function calls, memory spills, etc.). However, these references address only few locations and are not included in the FORAY model after being filtered out by the Step 4 of Algorithm 1. Note also that all functions that are called from different contexts are considered to be inlined (as explained in Section 4) for the purpose of calculating the number of loops and references in our experiments.

In summary, memory references that can be represented in FORAY form and are not in this form in a source code account for a large portion of memory activity in typical embedded applications. Our FORAY-GEN approach helps to automate the extraction of such references and thus expands the reach of SPM optimizations.

## 6. Conclusion

In this paper we introduce the notion of the FORAY model of a program which is another program consisting of *for* loops and array references with affine index expressions that models the memory behavior of the original program and that is amenable to optimizations. We presented FORAY-GEN, a technique for fully automated extraction of the FORAY model from a source program. We show how this model can be used to expand the reach of contemporary memory optimization techniques, increasing the number of analyzable references two times on average for typical embedded multimedia benchmarks. Our FORAY-GEN approach thus significantly increases the reach of SPM optimizations while freeing designers from the cumbersome and error-prone task of manual code analysis, thereby improving designer productivity. Our future work will study the interdependency of the FORAY models on the input data set used for profiling.